\documentclass[prd,singlecolumn,superscriptaddress,showpacs,preprint,12pt, nofootinbib]{revtex4-1}
\usepackage{comment}
\usepackage{amsmath,amssymb,esint,ulem}
\usepackage{verbatim}
\usepackage{graphicx}
\usepackage{mathrsfs}  

\usepackage{ stmaryrd }

\usepackage[
      colorlinks=true,
      linkcolor=blue,
      urlcolor=magenta,
      filecolor=green,
      citecolor=red,
      pdfstartview=FitV,
      pdftitle={},
        pdfauthor={Daniel Grumiller, Alfredo Perez, M.M. Sheikh-Jabbari, Ricardo Troncoso and Celine Zwikel},
        pdfsubject={},
        pdfkeywords={},
        pdfpagemode=None,
        bookmarksopen=true
      ]{hyperref}
\usepackage{color}
\renewcommand{\emph}[1]{{\it #1}}

\DeclareFontFamily{OT1}{rsfs}{}
\DeclareFontShape{OT1}{rsfs}{m}{n}{ <-7> rsfs5 <7-10> rsfs7 <10->rsfs10}{} 
\DeclareMathAlphabet{\mycal}{OT1}{rsfs}{m}{n}

\newcommand{\bel}[1]{ \begin{equation}\label{#1} }
\newcommand{\ee}{\end{equation}}
\newcommand{\bea}[1]{\begin{eqnarray}\label{#1} }
\newcommand{\eea}{\end{eqnarray}}
\newcommand{\p}{\partial}

\makeatletter \@addtoreset{equation}{section}

\newcommand{\eq}[2]{\begin{equation} #1 \label{#2} \end{equation}}

\DeclareMathOperator{\extdm}{d}

\newcommand{\extd}{\extdm \!}
\newcommand{\bcm}{}

\newcommand{\be}{\begin{equation}}
\newcommand{\bi}{\begin{itemize}}
\newcommand{\ei}{\end{itemize}}
\newcommand{\bt}{\begin{tabular}}
\newcommand{\et}{\end{tabular}}
\newcommand{\bc}{\begin{center}}
\newcommand{\ec}{\end{center}}

\def\tH{{\textrm{\tiny H}}}

\def\one{{\hbox{ 1\kern-.8mm l}}}
\newcommand{\Dslash}{\not{\hbox{\kern-4pt $D$}}}
\newcommand{\pdslash}{\not{\hbox{\kern-2pt $\partial$}}}

\newcommand{\ba}{\begin{array}}
\newcommand{\ea}{\end{array}}

\def\bbox{{\,\lower0.9pt\vbox{\hrule \hbox{\vrule height 0.2 cm
\hskip 0.2 cm \vrule height 0.2 cm}\hrule}\,}}
\newcommand{\dsl}{\pa \kern-0.5em /}

\newcommand{\vp}{\varphi}

\newcommand\snote[1]{\textcolor{blue}{\bf [Sh:\,#1]}}

\newcommand{\rindler}{\kappa} 
\newcommand{\area}{\Omega} 
\newcommand{\bcpt}{\eta} 
\newcommand{\sutr}{{\cal P}} 
\newcommand{\suro}{{\cal J}} 
\newcommand{\cofa}{\Phi} 

\begin{document}

\newcommand{\mytitle}{Spacetime structure near generic horizons and soft hair}

\title{\Large{\mytitle}}

\author{Daniel Grumiller}
\email{grumil@hep.itp.tuwien.ac.at}
\affiliation{Institute for Theoretical Physics, TU Wien, Wiedner Hauptstrasse 8--10/136, A-1040 Vienna, Austria}
\affiliation{Centro de Estudios Cient\'ificos (CECs), Av.~Arturo Prat 514, Valdivia, Chile}
\affiliation{School of physics, Inst.~for research in fundamental sciences (IPM), P.O.Box 19395-5531, Tehran, Iran}

\author{Alfredo P\'{e}rez}
\email{aperez@cecs.cl}
\affiliation{Centro de Estudios Cient\'ificos (CECs), Av.~Arturo Prat 514, Valdivia, Chile}

\author{M.M.~Sheikh-Jabbari}
\email{jabbari@theory.ipm.ac.ir}
\affiliation{School of physics, Inst.~for research in fundamental sciences (IPM), P.O.Box 19395-5531, Tehran, Iran}
\affiliation{The Abdus Salam ICTP, Strada Costiera 11, 34151 Trieste, Italy}

\author{Ricardo Troncoso}
\email{troncoso@cecs.cl}
\affiliation{Centro de Estudios Cient\'ificos (CECs), Av.~Arturo Prat 514, Valdivia, Chile}

\author{C\'eline Zwikel}
\email{zwikel@hep.itp.tuwien.ac.at}
\affiliation{Institute for Theoretical Physics, TU Wien, Wiedner Hauptstrasse 8--10/136, A-1040 Vienna, Austria}

\date{\today}

\preprint{CECS-PHY-18/01, IPM/P-2019/009, TUW--18--03}

\begin{abstract} 

We explore the spacetime structure near non-extremal horizons in any spacetime dimension greater than two and discover a wealth of novel results: 1.~Different boundary conditions are specified by a functional of the dynamical variables, describing inequivalent interactions {at the horizon} with a thermal bath. 2.~The near horizon algebra of a set of boundary conditions, labeled by a parameter $s$,  
is given by the semi-direct sum of diffeomorphisms at the horizon with ``spin-$s$ supertranslations''. For $s=1$ we obtain the first explicit near horizon realization of the Bondi--Metzner--Sachs algebra. 3.~For another choice, we find a non-linear extension of the Heisenberg algebra, generalizing recent results in three spacetime dimensions.  This algebra allows to recover the aforementioned (linear) ones as composites. 4.~These examples allow to equip not only black holes, but also cosmological horizons with soft hair.  We also discuss implications of soft hair for black hole thermodynamics and entropy.
  \end{abstract}


\maketitle

\tableofcontents
\section{Introduction} 
Horizons are among the most remarkable entities in geometries with Lorentzian signature, appear in different contexts and lead to rich phenomenology, see e.g.~\cite{Bardeen:1973gs,Bousso:2002ju,Padmanabhan:2003gd,Ashtekar:2004cn} and refs.~therein. Black holes, by their very definition, have horizons \cite{Hawking:1973}, which are in turn responsible for their peculiar classical properties that are reflected in observables such as X-ray spectra from accretion disks \cite{Shakura:1972te,Remillard:2006fc}, the gravitational wave spectrum of black hole mergers \cite{Blanchet:1995ez,Pretorius:2005gq,Blanchet:2006zz,Abbott:2016blz}, or their shadow, as observed recently by the Event Horizon Telescope \cite{EHT:2019}. Moreover, horizons are responsible for semi-classical properties, such as Hawking temperature and black hole evaporation \cite{Hawking:1974sw} or the universal result for the Bekenstein--Hawking entropy \cite{Bekenstein:1973ur}. Cosmological horizons also have thermal properties \cite{Gibbons:1977mu} and the very predictions of inflation \cite{Starobinsky:1979ty,Guth:1980zm,Linde:1981mu} about power spectra of primordial perturbations is a result of horizon-crossing of the quantum fluctuations \cite{Mukhanov:1981xt,Guth:1982ec,Mukhanov:1990me}. There are observer dependent acceleration horizons that lead to Rindler quanta \cite{Rindler:1966zz,Birrell:1982ix} and the Unruh effect \cite{Unruh:1976db}.  Horizons also feature prominently in analog black holes (dumb holes) \cite{Unruh:1980cg, Barcelo:2005fc}. 

Thermodynamical properties of horizons have led to the idea of deriving Einstein's equations (or generalizations thereof) from thermodynamics \cite{Jacobson:1995ab,Padmanabhan:2009vy,Verlinde:2010hp}. Understanding horizons and the associated microstates quantum mechanically is an ongoing research endeavor where considerable progress has been made already \cite{Strominger:1996sh,Strominger:1997eq,Guica:2008mu,VanRaamsdonk:2010pw, Maldacena:2013xja,Harlow:2014yka}, partly thanks to the holographic principle \cite{'tHooft:1993gx,Susskind:1995vu} as manifested by the AdS/CFT correspondence \cite{Maldacena:1997re,Gubser:1998bc,Witten:1998qj}. 

Nevertheless, non-extremal horizons (i.e., horizons at finite temperature) remain largely mysterious  \cite{Hawking:1976ra,Preskill:1992tc,Mathur:2005zp,Skenderis:2008qn,Almheiri:2012rt,Papadodimas:2013wnh}. No reliable and universal results for their microstates are known so far. Even in the simpler case of gravity in three spacetime dimensions concrete microstate proposals such as horizon fluff  \cite{Afshar:2016uax,Sheikh-Jabbari:2016npa} require ad-hoc assumptions \cite{Afshar:2017okz}. It is thus of considerable interest to understand non-extremal horizons as deeply as possible. 

Another motivation for our work is to clarify possible relations between various different proposals for infinite-dimensional near horizon symmetries (including symmetries resembling Bondi--Metzner--van~der~Burg--Sachs (BMS) \cite{Bondi:1962,Sachs:1962}, Virasoro and others), which may seemingly appear to be in conflict with each other, see e.g.~\cite{Donnay:2015abr,Afshar:2016wfy,Carlip:2017xne,Penna:2017bdn,Haco:2018ske} and also \cite{tHooft:1990fkf,tHooft:1991uqr,Susskind:1993if,Carlip:1994gy,Balachandran:1994up,Carlip:1995cd,Ashtekar:1997yu,Carlip:1998wz,Hotta:2000gx,tHooft:2006xjp,Majhi:2012tf,Afshar:2015wjm,Donnay:2016ejv,Lust:2017gez,Penna:2018gfx,Haco:2019ggi,Donnay:2019zif,Donnay:2019jiz,Chandrasekaran:2019ewn}.  

In this work we derive generic properties of non-extremal horizons, assumed to be in equilibrium with a thermal bath, in any spacetime dimension greater than two. The physical properties of the thermal bath are modelled by the way we impose boundary conditions, and we shall describe various different well-motivated choices leading to infinite-dimensional near horizon symmetries. We prove that they generically span soft hair excitations in the sense of Hawking, Perry and Strominger \cite{Hawking:2016msc} (see also \cite{Averin:2016ybl,Compere:2016hzt,Mirbabayi:2016axw,Hawking:2016sgy,Carney:2017jut,Bousso:2017dny,Donnelly:2017jcd,Angelopoulos:2018yvt,Raposo:2018xkf}). While our methods are general, we focus on Einstein gravity (possibly with cosmological constant).

We discuss in detail a wide class of boundary conditions, labelled by a parameter $s$, whose near horizon symmetry algebra is given by the semi-direct sum of diffeomorphisms at the spacelike section of the horizon with spin-$s$ supertranslations. 
In particular, the case of $s=1$ contains the first explicit near horizon realization of BMS in any dimension.  Moreover, we also propose a special type of boundary conditions whose near horizon symmetries are described by a suitable non-linear extension of the Heisenberg algebra, generalizing the results found in three spacetime dimensions \cite{Afshar:2016wfy,Afshar:2016kjj}. This algebra allows to recover the aforementioned ones as composites, in a way that we shall make precise.

Since our analysis is valid for any non-extremal horizon it applies in particular to generic Kerr black holes \cite{Kerr:1963ud} (including Schwarzschild) and its generalizations that include NUT charge \cite{Taub:1950ez,Newman:1963yy}. 

\section{Near horizon expansion} Let us consider a metric in $D\geq 3$ spacetime dimensions that features a non-extremal horizon (see \cite{Hawking:1973,Hayward:1993wb,Ashtekar:2000sz,Ashtekar:2002ag} and refs.~therein for various notions of horizon). Our working definition of a non-extremal horizon is that the line-element can be brought into a Rindler-like form \cite{Rindler:1966zz},
\eq{
\extd s^2 = -\rindler^2\,\rho^2\,\extd t^2 + \extd\rho^2 + \area_{ab}\,\extd x^a\extd x^b + \dots
}{eq:nh2}
where $\rindler$ is surface gravity (with $\rindler\neq 0$ to guarantee non-extremality), $t$ is time, and $\rho$ is some radial coordinate that vanishes at the horizon. The ``horizon metric'', $\Omega_{ab}$ (where $a=1,2,\dots(D-2)$), has non-vanishing determinant,  $\area=\det\area_{ab}\neq0$, to avoid degeneracy. The ellipsis denotes higher order terms or rotation/boost terms, which we spell out explicitly by defining a suitable near horizon expansion for the metric. 

Our main assumption is that the horizon is free from singularities and the metric permits a Taylor expansion in $\rho$ in the near horizon region. Consistency with our assumptions above and with smoothness of the horizon determines the near horizon behavior of the metric, which in a co-rotating frame is given by~\footnote{Additional terms omitted in \eqref{eq:nh1} are either pure gauge or inconsistent with integrability of boundary charges or refer to a rotating frame. Non-integrable charges would be needed if we were interested in dynamical situations where gravitational waves (or matter) have a net flux through the horizon. In our current work we exclude this possibility, {while allowing for cases with equal in and out fluxes through the horizon, like the case of large AdS black holes which are in equilibrium with their own Hawking radiation}.}
\begin{align}
g_{tt}& =-\kappa^2 \rho^2 + {\cal O}(\rho^3) & g_{\rho\rho} & = 1 + {\cal O}(\rho) \nonumber \\
g_{t\rho}& ={\cal O}(\rho^2)
&  g_{\rho a}  &= f_{\rho a}\, \rho + {\cal O}(\rho^2) 
\label{eq:nh1} \\
g_{ta}&= f_{ta}\,\rho^2 + {\cal O}(\rho^3)  & 
 g_{ab} &={\hat\area_{ab}} + {\cal O}(\rho^2) \nonumber\,.
\end{align}  
{where $\hat\Omega_{ab}$ and $\Omega_{ab}$ are diffeomorphic to each other.} 
All coefficients in \eqref{eq:nh1} can depend on time $t$ and the transverse coordinates $x^a$, but not on the radius $\rho$. 

The near horizon expansion \eqref{eq:nh1} is preserved by a set of diffeomorphisms generated by vector fields $\xi^\mu$
\eq{
\xi^t = \frac{\bcpt}{\kappa} + {\cal O}(\rho),\qquad 
\xi^\rho = {\cal O}(\rho^2),\qquad 
\xi^a = \bcpt^a + {\cal O}(\rho^2)
}{eq:diffeos2}
where $\bcpt^a$ depends arbitrarily on $x^a$, while $\bcpt$ depends additionally on $t$ subject to the condition $\partial_t\bcpt+\bcpt^a\partial_a\kappa=\delta\kappa$. The dynamical fields, $\sutr$ and $\suro_a$, defined by
\eq{
\sutr :=\frac{\sqrt{\Omega}}{8\pi G}  \qquad\quad
\suro_a:=\frac{\sqrt{\Omega}}{16\pi G \kappa} \, \Big(\partial_t f_{\rho a} - 2 f_{ta} \Big) 
}{eq:nh4}
transform as
\begin{subequations}
\label{eq:nh5}
\begin{align}
    \delta\sutr &= \bcpt^a\partial_a\sutr + \sutr\partial_a\bcpt^a \\
    \delta\suro_a &= \sutr\partial_a\bcpt + \bcpt^c\partial_c\suro_a  + \suro_c\partial_a\bcpt^c + \suro_a\partial_c\bcpt^c\,.
\end{align}
\end{subequations}

The  near horizon symmetries are spanned by generators{, co-dimension two surface charges,} that can be calculated using different methods, see e.g.~\cite{Regge:1974zd,Wald:1999wa,Barnich:2001jy}. Their variations
\eq{
\delta Q[\bcpt,\,\bcpt^a]=\int\extd^{D-2}x\, \big[\bcpt\,\delta\sutr + \bcpt^a\,\delta\suro_a\big]
}{eq:nh3}
turn out to be non-trivial and finite. To unveil the near horizon symmetries as next step we impose boundary conditions that lead to integrable charges.

\section{Fixing boundary conditions} To obtain the charges $Q[\bcpt,\,\bcpt^a]$ (rather than their variations) we need precise boundary conditions by specifying the allowed variations of the metric. To admit the most general boundary conditions we rewrite the near horizon expansion \eqref{eq:nh1} in a rotating frame,  implemented through an Arnowitt--Deser--Misner decomposition \cite{Arnowitt:1962hi}. Metric components $g_{t\mu}$ (with $\mu=t,\rho,a$) determine lapse function $N$ and shift vector $N^i=(N^\rho,N^a)$, which are Lagrange multipliers in the canonical formulation. Their leading order boundary behavior determines the sources (in holographic literature) or, equivalently, the chemical potentials (in black hole thermodynamics literature). In a rotating frame, lapse and shift expand as
\eq{
N = \mathcal{N}\rho + {\cal O}(\rho^2),\qquad   N^a = \mathcal{N}^a + {\cal O}(\rho^2),\qquad N^\rho = {\cal O}(\rho^2).
}{eq:kerr155}
The co-rotating fall-off is recovered for $\mathcal{N}^a=0$, $\mathcal{N}=\kappa$. 

Chemical potentials $\mathcal{N}$ and $\mathcal{N}^a$ appear in the boundary term $I_B$ that has to be added to the bulk Hamiltonian action for a well-defined variation principle \cite{Regge:1974zd}. It is evaluated in the limit of small $\rho$ and its variation reads
\eq{
\delta I_B = -\int\extd t\extd^{D-2}x\,\big(\mathcal{N}\,\delta\sutr + \mathcal{N}^a\,{\delta\mathcal{J}_a}\big)
}{eq:kerr156}
where {$\mathcal{J}_a$} acquires additional terms as compared to \eqref{eq:nh4} in a rotating frame \footnote{In general, ${\cal J}_{a}=16 \pi G\Omega_{ab}\pi_{\left(0\right)}^{\rho b}$, where $\pi_{\left(0\right)}^{\rho b}$ stands for the leading term of the canonical momenta conjugate to $g_{\rho b}$.}. Following \cite{Perez:2016vqo}, integrability of the boundary term \eqref{eq:kerr156} generically requires
\eq{
\mathcal{N}=\frac{\delta { F}}{\delta\mathcal{P}}\qquad\qquad\mathcal{N}^{a}=\frac{\delta {F}}{\delta\mathcal{J}_{a}}
}{eq:nh2000} 
for some functional $F[\sutr,\,\mathcal{J}_a]=\int\extd^{D-2}x\;\mathcal{F}(\sutr,\,\mathcal{J}_a)$. Therefore, the boundary conditions are fully fixed only once this functional is specified.

As spacetime does not possess a boundary near the horizon, different choices for the chemical potentials \eqref{eq:nh2000} amount to different prescriptions for the black hole interactions with a thermal bath, without the need of specifying the microscopic structure of the surrounding atmosphere. Thus, we are following an old and standard practice, like in electromagnetism of different media: all of the relevant information about the microscopic structure of the material is summarized in Maxwell's theory through the precise way one fixes the chemical potential at the boundary of the body, e.g.~by fixing the time-component of the gauge-potential, $A_t$, for a conductor (Dirichlet), or its normal derivative for a dielectric medium (Neumann).

Compatibility of time-evolution with near horizon symmetries implies that the symmetry generator parameters in \eqref{eq:nh3} acquire the same field-dependence as the chemical potentials in \eqref{eq:nh2000}, with $\mathcal{N}\to\bcpt$ and $\mathcal{N}^a\to\bcpt^a$. Thus, integrability of the boundary term \eqref{eq:kerr156} implies integrability of the symmetry generators \eqref{eq:nh3}.

We focus next on special choices for $\mathcal{F}(\sutr,\,\suro_a)$ that lead to different sets of boundary conditions with an infinite number of near horizon symmetries.

\section{BMS-like symmetries} Retaining the full set of diffeomorphisms along the spacelike section of the horizon leads to an infinite amount of near horizon symmetries. This is achieved e.g.~by fixing $\delta \mathcal{N}^{a}=0$, while the lapse $\mathcal{N}$ can still be allowed to depend on $\mathcal{P}$. We propose a family of boundary conditions labeled by a parameter $s$, 
\eq{
\mathcal{F}(\sutr,\,\suro_a)=\mathcal{N}_{(s)} \, \frac{\sutr^{r+1}}{r+1}+\mathcal{N}^a \suro_a
}{eq:nh1999}
where $\mathcal{N}_{(s)}$ is fixed, $\delta \mathcal{N}_{(s)}=0$, and $r=s/(D-2)$. This choice, together with \eqref{eq:nh2000}, implies
\begin{equation}
\mathcal{N}=\mathcal{N}_{(s)}\mathcal{P}^{r}\qquad \Rightarrow \qquad \eta=\eta_{(s)}\mathcal{P}^{r}\,.\label{chempotBMS}
\end{equation}

The variation of the charges \eqref{eq:nh3} integrates as 
\eq{
Q[\bcpt_{(s)},\,\bcpt^a] = \int\extd^{D-2}x\,\big[\eta_{(s)}\mathcal{P}_{(s)} + \bcpt^a\suro_a\big]
}{eq:nh8}
with $\sutr_{(s)} = \sutr^{r+1}/(r+1)$. Hence, from \eqref{eq:nh5}, the transformation law of the fields is given by
\begin{subequations}
\label{eq:nh9}
\begin{align}
\delta\sutr_{(s)} &= \bcpt^a\,\partial_a\sutr_{(s)} +
(r+1)\,\sutr_{(s)}\,\partial_a\bcpt^a \\
\delta\suro_a &= (r+1)\, \sutr_{(s)}\,\partial_a\bcpt_{(s)} + r\,\bcpt_{(s)} \partial_a\sutr_{(s)} + \bcpt^c\partial_{c} \suro_a + \suro_a\partial_c\bcpt^c + \suro_c\partial_a\bcpt^c
\end{align}
\end{subequations}
so that the near horizon Poisson bracket algebra reads
\begin{align}
&\{\suro_a(x),\sutr_{\tiny{(s)}}(y)\} = \Big(r\sutr_{\tiny{(s)}}(y)\frac{\partial}{\partial x^a} - \sutr_{\tiny{(s)}}(x)\frac{\partial}{\partial y^a}\Big) \delta(x-y) \nonumber \\
&\{\sutr_{\tiny{(s)}}(x),\sutr_{(s)}(y)\}=0
\label{eq:bms} \\
&\{\suro_a(x),\suro_b(y)\} = \Big(\suro_a(y)\frac{\partial}{\partial x^b}-\suro_b(x)\frac{\partial}{\partial y^a }\Big) \delta(x-y) \nonumber
\end{align}

The algebra \eqref{eq:bms} is the semi-direct sum of diffeomorphisms at the $(D-2)$--dimensional spacelike section of the horizon, generated by $\suro_a(x)$, and a generalization of supertranslations spanned by $\sutr_{(s)}(x)$.

If the spacelike section of the horizon has the topology of $S^{D-2}$, the largest finite subalgebra is given by the semidirect sum of $so(D-1,1)$ (Lorentz) and ``spin-$s$ translations'', spanned by suitable subsets of $\suro_a(x)$ and $\sutr_{(s)}(x)$, respectively.

For $D=3$, the algebra \eqref{eq:bms} for $s=0$ agrees with the one found in \cite{Donnay:2015abr,Donnay:2016ejv} while for $s=1$ it is the BMS$_3$ algebra \cite{Ashtekar:1996cd,Barnich:2006av}. For generic $s$ it is the $W(0,-s)$ algebra \cite{Gao:2011,Parsa:2018kys}, where the supertranslations are generators with conformal weight $h=s+1$. 

For $D\geq 4$, if the horizon metric is restricted to be conformal to the round sphere, $\Omega_{ab}=\cofa^{2}\,\Omega_{ab}^{S^{D-2}}$, the parameters $\eta^{a}$ reduce to the conformal Killing vectors of ${S^{D-2}}$, so that its associated generators $\mathcal{J}_a$ exactly span the Lorentz algebra $so(D-1,1)$. It is then worth highlighting that our boundary conditions for $s=1$ provide the first explicit realization of the BMS$_D$ algebra 
as near horizon symmetries in four and higher dimensions. Further aspects of BMS$_D$ and its higher spin extensions are discussed in the Appendix \ref{Appen-1}. 

For $s=0$, the algebra carries ``scalar supertranslations'' and agrees with the result in \cite{Donnay:2015abr,Donnay:2016ejv} for $D=3,4$.

\section{Heisenberg-like symmetries} The parameters $\bcpt^a$ in \eqref{eq:nh3} are vectors, while their corresponding charges are 
1-form densities of weight one. It is natural to swap their role, in the sense that the charges are 1-forms and their corresponding parameters become densities. This set of boundary conditions is obtained by choosing
\eq{
    \mathcal{F}(\sutr,\,\suro_a)=\mathcal{N}_{\textrm{\tiny H}} \mathcal{P} +\mathcal{N}^{a}_{\textrm{\tiny H}}\,\mathcal{J}_{a}\,\mathcal{P}^{-1}
}{eq:whatever}
with $\mathcal{N}_{\textrm{\tiny{H}}}$ and $\mathcal{N}^{a}_{\textrm{\tiny{H}}}$ fixed $(\delta \mathcal{N}_{\textrm{\tiny{H}}}=0,\; \delta \mathcal{N}^{a}_{\textrm{\tiny{H}}}=0)$. The chemical potentials follow from \eqref{eq:nh2000}, so that the symmetry parameters in \eqref{eq:nh3} are now given by
\eq{
    \eta^a=\eta_{\textrm{\tiny{H}}}^{a}\,\mathcal{P}^{-1}\qquad\qquad\eta=\eta_{\textrm{\tiny{H}}}-{\eta_{\textrm{\tiny{H}}}^{a}\,\suro_a\,\mathcal{P}^{-2}}\,.
}{eq:nolabel}
The variation of the generators \eqref{eq:nh3} then integrates as 
\eq{
Q_{\textrm{\tiny{H}}}[\bcpt_{\textrm{\tiny{H}}},\,\bcpt^a_{\textrm{\tiny{H}}}]=\int\extd^{D-2}x\,\big[\bcpt_{\textrm{\tiny{H}}}\,\sutr + \bcpt^a_{\textrm{\tiny{H}}}\,\suro_a^{\textrm{\tiny H}}\big]\,
}{eq:angelinajolie}
with the 1-form  $\suro_a^{\textrm{\tiny H}}:=\suro_a \mathcal{P}^{-1}$. 
The transformation laws \eqref{eq:nh5} now yields
\begin{subequations}
\label{eq:nh11}
\begin{align}
    \delta\sutr &= \partial_a\bcpt^a_{\textrm{\tiny H}} \label{eq:nh11a}\\
    \delta\suro_a^{\textrm{\tiny H}} &=  \partial_a\bcpt_{\textrm{\tiny H}} - \bcpt^b_{\textrm{\tiny H}}\,F_{ab}\sutr^{-1}\label{eq:nh11b}
\end{align}
\end{subequations}
where $F_{ab}:=\partial_a\suro_b^{\textrm{\tiny{H}}}-\partial_b\suro_a^{\textrm{\tiny{H}}}$. The
transformation laws  \eqref{eq:nh11} establish the Poisson bracket algebra
\begin{subequations}
\label{alg}
\begin{align}
\left\{\mathcal{J}_{a}^{\textrm{\tiny H}}\left(x\right),\,\mathcal{P}\left(y\right)\right\}  & = \frac{\partial}{\partial x^{a}}\delta\left(x-y\right) \label{alg1}\\
\left\{ \mathcal{P}\left(x\right),\,\mathcal{P}\left(y\right)\right\}  & =0\\
\left\{ \mathcal{J}_{a}^{\textrm{\tiny H}}\left(x\right),\,\mathcal{J}_{b}^{\textrm{\tiny H}}\left(y\right)\right\}  & 
=\mathcal{P}^{-1}(x) F_{ba}\left(x\right)\delta\left(x-y\right)\,.\label{H-alg-c}
\end{align}
\end{subequations}
We note that \eqref{alg1} implies $\{\mathcal{P}(x),\,F_{ab}\left(y\right)\}=0$.

In three dimensions $F_{ab}$ identically vanishes, so that the boundary conditions reduce to those in \cite{Afshar:2016wfy,Afshar:2016kjj}, which accommodate ``soft hairy'' black hole solutions, while \eqref{alg} becomes equivalent to two copies of $\hat{u}(1)$ current algebras.
 
In $D>3$ dimensions the phase space restricted to configurations with  $F_{ab}=0$ is preserved under the full set of asymptotic symmetries in \eqref{eq:nh11}. Examples are toroidal Kerr-AdS black holes and Schwarzschild black holes, together with their soft hair excitations. In these cases the 1-form $\suro_a^{\textrm{\tiny{H}}}$ is locally exact, $\suro_a^{\textrm{\tiny H}}=:8\pi G\partial_a{\cal Q}$. Then, the only non-vanishing Poisson bracket \eqref{alg1},
\begin{equation}\label{eq:Heisenberg}
\{{\cal{Q}}(x),\,\sutr(y)\}=\frac{1}{8\pi G}\,\delta(x-y)
\end{equation}
yields the Heisenberg algebra. 
Quantizing the Poisson bracket algebra \eqref{eq:Heisenberg} (viz., replacing $i\{,\}$ by commutators $[,]$) yields canonical commutation relations with the factor $1/(4G)$ playing the role of Planck's constant $h$ \cite{Grumiller:2018scv}.  This simple result could be relevant in semi-classical descriptions of black holes.~\footnote{Symmetry generating diffeomorphisms that lead to the Heisenberg algebra \eqref{eq:Heisenberg} consist of  $x^a$-dependent time translations $\sutr (x)$  and area preserving shear deformations of the horizon $\suro_{a}^{\textrm{\tiny H}}(x)$. The latter are crucial in the membrane paradigm \cite{Thorne:1986iy,Price:1986yy}. Our analysis can provide the setting to formulate and quantize the membrane paradigm \cite{Grumiller:2018scv}.}

\section{Compositeness, soft hair and thermodynamics} 
{We discussed two classes of algebras, the BMS-like ones \eqref{eq:bms} in which the right-hand-side (RHS) of commutators are linear in the generators $\left(\mathcal{J}_a,\mathcal{P}_{(s)}\right)$ and the Heisenberg-like algebra spanned by $\left(\suro_a^{\textrm{\tiny H}},\mathcal{P}\right)$, in which  the RHS of (\ref{H-alg-c})  are nonlinear in generator $\sutr$.} 
Remarkably, {generators of} the BMS-like algebras \eqref{eq:bms} emerge as composites in terms of {generators of} the Heisenberg-like algebra \eqref{alg}, $\mathcal{J}_a=\suro_a^{\textrm{\tiny{H}}}\sutr$, and $\sutr_{(s)}=\sutr^{r+1}/(r+1)$. In this sense, the Heisenberg-like {generators} ($\suro_a^{\textrm{\tiny H}},\sutr$) 
{are the} building blocks. We have thus generalized this feature observed first in three \cite{Afshar:2016kjj} to arbitrary dimensions. Note that in $D\geq 4$ the nonlinearity of the Heisenberg-like algebra is the key to establish the map between the {generators}.

All boundary conditions discussed here allow soft hair excitations in the sense of \cite{Hawking:2016msc}, i.e., gravitational excitations that carry no energy, but nonetheless are not pure gauge. For the BMS-like boundary conditions, the near horizon total Hamiltonian, given by the generator of unit time translations, in a non-rotating frame reads
$H_{\left(s\right)}=Q\left[\partial_{t}\right]=\int\extd^{D-2}x\,\mathcal{N}_{\left(s\right)}\mathcal{P}_{\left(s\right)}$. The spin-$s$
supertranslations $\mathcal{P}_{\left(s\right)}$ correspond to soft hair charges since they commute with $H_{\left(s\right)}$, see \eqref{eq:bms}. For Heisenberg-like boundary conditions, the {generators} $\mathcal{P}$ also stand for soft hair charges, since they commute with the near horizon Hamiltonian, which in a non-rotating frame reads
$H_{\textrm{\tiny H}}=Q\left[\partial_{t}\right]=\int\extd^{D-2}x\, 
\mathcal{N}_{\textrm{\tiny H}}\mathcal{P}$. 
  
We address now the Bekenstein--Hawking entropy. For BMS-like boundary conditions it reads 
$S=A/(4G)=2\pi\left(r+1\right)^{\frac{1}{r+1}}\int\extd^{D-2}x\,(\mathcal{P}_{\left(s\right)})^{\frac{1}{r+1}}$. Although soft hair excitations do not contribute to the energy,  for $s\neq 0$ they contribute to the entropy through the modes of $\mathcal{P}_{(s)}$. Only for $s=0$ or for the Heisenberg-like boundary conditions soft hair excitations do not contribute to the Bekenstein--Hawking entropy, which is given by the zero mode of $\mathcal{P}_0=\mathcal{P}$ \cite{Donnay:2015abr,Afshar:2016wfy}
\eq{
S = 2\pi\,\sutr_0
}{eq:entropy}
with $\sutr_0=\int\extd^{D-2}x\,\sutr$.

Different choices of boundary conditions generically describe  inequivalent thermodynamic ensembles. The chemical potentials correspond to the variables that are kept fixed. Demanding smoothness of the metric around the horizon implies that the lapse and shift are given by $\mathcal{N}=\kappa=2\pi\beta^{-1}$ and $\mathcal{N}^{a}=0$, where inverse temperature $\beta$ is the Euclidean time period. For the BMS-like boundary conditions the chemical potentials \eqref{chempotBMS} are fixed as $\mathcal{N}_{\left(s\right)}=\frac{2\pi}{\beta}\left[\left(r+1\right)\mathcal{P}_{\left(s\right)}\right]^{-\frac{r}{r+1}}$ and thus, the first law is fulfilled as expected $\delta S=\beta\,\delta H_{(s)}$, where the variation of the total Hamiltonian includes the internal energy, as well as work terms. 
For the Heisenberg-like boundary conditions, the chemical potentials are given by 
$\mathcal{N}_{\textrm{\tiny H}}=2\pi\beta^{-1}$, $\mathcal{N}^{a}_{\textrm{\tiny{H}}}=0$ and the first law reads $\delta S=\beta\,\delta H_{\textrm{\tiny{H}}}$. Note that in the latter case, as well as for $s=0$, temperature $T=\beta^{-1}$   is state-independent.

\section{Kerr black hole example} Our results can be applied to arbitrary non-extremal black holes in diverse dimensions, for instance black holes with cosmological constant or NUT charge. Here we give the essentials for the most interesting black hole, non-extremal Kerr \cite{Kerr:1963ud}. Since the BMS-like {generators} are composites of the Heisenberg-like ones, it is enough to perform the analysis in the latter case. {For this choice we have mixed boundary conditions that describe how the Kerr black hole interacts with a thermal bath: the metric component $g_{tt}$ is fixed, the metric component $g_{t\rho}$ is irrelevant and the
metric components $g_{ta}$ are allowed to fluctuate in a state-dependent way. Translated to Maxwell's theory these boundary conditions mean that the black hole horizon behaves like a conductor only with respect to
 $g_{tt}$, but not with respect to $g_{ta}$.}
The near horizon Heisenberg-like generators for Kerr black holes with event/inner horizon radii $r_\pm$ are given by ($\theta\in[0,\pi)$, $\varphi\sim\varphi+2\pi$)
\begin{align}\label{eq:kerr}
\sutr = \frac{r_+(r_+ + r_-)}{8\pi G}\,\sin\theta,\qquad 
    \suro_a^\tH = \delta_a^\vp\,r_-\,\frac{r_-(r_- - r_+)\cos^2\theta - r_+(3 r_+ + r_-)}{2\sqrt{r_+r_-}\,(r_+ + r_-\cos^2\theta)^2}\,\sin^2\theta  \,.
\end{align}
$\sutr$ only has a monopole contribution (in the sense that it is proportional to the 2-dimensional volume factor $\sin\theta$), while $\suro_a^\tH$  only have a coexact part, $\suro_a^\tH=\varepsilon_a{}^b\,\partial_b\psi$, where $\psi=2\arctan{U}+\frac{r_+-r_-}{r_++r_-}U$ with $U=\sqrt{{r_-}/{r_+}}\cos\theta$. The field strength defined below \eqref{eq:nh11b} is non-zero, $F_{ab}\neq 0$, but the flux integrated over the horizon vanishes,  $\int_H F=0$. Further details and a generalization to Kerr black holes with NUT charges (which introduces fluxes $\int_H F\neq0$) can be found in the Appendix \ref{Appen-2}.

\section{Comments and further developments}
In the present work we focused on non-extremal black hole horizons. Our results also apply to cosmological horizons. Indeed, $\rho=0$  in \eqref{eq:nh1} could be the observer horizon associated with a cosmological de~Sitter patch. It is worthwhile to apply our analysis to cosmology and study possible consequences for cosmic perturbation theory and hence cosmological observables with soft hair. Our near horizon charges could be relevant in the derivation of cosmological consistency relations \cite{Hinterbichler:2012nm, Joyce:2014aqa} or for the infinite set of Ward identities of the adiabatic modes \cite{Hinterbichler:2013dpa}.


An interesting generalization is to include matter or gravitational wave fluxes through the horizon, allowing a wider class of configurations, including non-stationary black holes. Such an analysis may pave the way to address Hawking radiation and the information paradox. 

While we focused on matterless Einstein gravity, results from three dimensions \cite{Afshar:2016wfy,Setare:2016jba,Grumiller:2016kcp,Afshar:2016kjj,Setare:2016vhy,Ammon:2017vwt,Grumiller:2017jft,Grumiller:2017otl,Setare:2017xlu}
suggest that the BMS-like and Heisenberg-like algebras are universal and apply also to General Relativity with matter, higher derivative theories and possibly other modifications of Einstein gravity. It could be rewarding to verify (or falsify) this claim by considering such examples.

  
\vspace{3mm}

\noindent{\textbf{Acknowledgments}}

We are grateful to Hamid Afshar, Dario Francia, Oscar Fuentealba, Hern\'an Gonz\'alez, Wout Merbis,
 Jorge Nore\~na, Emilio Ojeda, Miguel Pino, Pablo Rodr\'iguez, David Tempo, Raphaela
 Wutte and Hossein Yavartanoo for collaboration on aspects of near horizon symmetries.

DG was supported by the Austrian Science Fund (FWF), projects P~28751, P~30822.
MMShJ would like to thank the hospitality of ICTP HECAP where this work finished and acknowledges the support by 
INSF grant No 950124 and Saramadan grant No. ISEF/M/98204.
CZ was supported by the Austrian Science Fund (FWF), projects P 30822 and M 2665. 
DG and MMShJ acknowledge the Iran-Austria IMPULSE project grant, supported and run by Khawrizmi University.
DG, AP and RT acknowledge travel support from the Conicyt grant REDES170052. 
We also thank the hospitality of the Erwin Schr\"odinger Institute (ESI) in Vienna during the program
  \href{http://quark.itp.tuwien.ac.at/~grumil/ESI2019/}{`Higher Spins and 
  Holography'}, as well as the organizers of the ``Second Hermann Minkowski Meeting on the Foundations of Spacetime Physics'' in Albena, where part of this work was completed. This research has been partially supported by Fondecyt grants No 1161311, 1171162, 1181031 and 1181496. The Centro de Estudios Cient\'ificos (CECs) is funded by the Chilean Government through the Centers of Excellence Base Financing Program of Conicyt.

\appendix


\setcounter{equation}{0}

\section{More on \texorpdfstring{BMS$\boldsymbol{^{(s)}_D}$}{BMSD} algebras}\label{Appen-1}

In this section we discuss some details of the BMS-like algebra in \eqref{eq:bms}, possessing ``spin-$s$ supertranslations'' in $D$ spacetime dimensions, which we denote by BMS$^{(s)}_D$. 

\subsection{Three spacetime dimensions} In $D=3$ the spacelike section of the horizon is topologically $S^1$, with angular coordinate denoted by $\varphi\sim\varphi+2\pi$. Expanding the generators $\sutr^{(s)}$ and $\suro$ (which in $D=3$ carry no indices) in Fourier modes
\eq{
\sutr^{(s)}_n = \frac{1}{2\pi}\,\oint\extd\varphi\,\sutr^{(s)}\,e^{in\varphi}, \qquad 
\suro_n =  \frac{1}{2\pi}\,\oint\extd\varphi\,\suro\,e^{in\varphi}
}{eq:fi1}
the algebra \eqref{eq:bms} reads
\begin{align}
    i\{\suro_n,\,\sutr^{(s)}_m\} &= (sn-m)\,\sutr_{n+m}^{(s)} \nonumber \\
    i\{\sutr^{(s)}_n,\,\sutr^{(s)}_m\} &= 0 \label{eq:fi2}\\
    i\{\suro_n,\,\suro_m\} &= (n-m)\,\suro_{n+m}\,. \nonumber
\end{align}
The last bracket describes the Witt algebra, corresponding to diffeomorphisms of $S^1$; and thus, the first bracket shows that $\sutr^{(s)}$ behaves as a primary field of conformal weight $h=s+1$ (see e.g. \cite{diFrancesco}). The middle bracket then shows that $\sutr^{(s)}_n$ generate an infinite set of mutually commuting diffeomorphisms, describing spin-$s$ supertranslations. 

For $s=1$ the algebra \eqref{eq:fi2} coincides with the standard (non-centrally extended) BMS$_3$ algebra \cite{Ashtekar:1996cd,Barnich:2006av}, while for $s=0$ the algebra \eqref{eq:fi2} reduces to the one in \cite{Donnay:2015abr}. For generic $s$, the algebra \eqref{eq:fi2} is $W(0,-s)$ \cite{Gao:2011,Parsa:2018kys} which may be obtained as algebraic deformation of BMS$_3$ \cite{Parsa:2018kys}.

It is worth highlighting that for our boundary conditions, higher spin supertranslations naturally arise in the near horizon behavior of Einstein gravity in vacuum, without the need of introducing higher spin fields (see e.g. \cite{Afshar:2013vka,Gonzalez:2013oaa,Gary:2014ppa,Matulich:2014hea})

\subsection{Four spacetime dimensions} In $D=4$, the subset of the algebra BMS$^{(s)}_4$ \eqref{eq:bms} spanned by $\mathcal{J}_a$ consists of generic diffeomorphisms in two dimensions. Thus, for $s=1$, our near horizon algebra BMS$^{(1)}_4$ agrees with the algebra reported in \cite{Campiglia:2014yka}, obtained through a different analysis at null infinity.

As mentioned in the main text, if the horizon metric is assumed to be conformal to the sphere $S^{2}$, the subset of globally well-defined diffeomorphisms at the spacelike section of the horizon breaks down to $SO(3,1)$. Nonetheless, as in \cite{Barnich:2011mi}, if the conformal Killing vectors of $S^2$ are allowed to be only locally defined, the Lorentz group becomes enlarged to the full set of ``superrotations'' spanned by two copies of the Witt algebra; at the expense of having divergent charges, but finite associated generators $\mathcal{J}_a$ (except for isolated points). With these assumptions, the mode expansion of the BMS$^{(s)}_4$ algebra in \eqref{eq:bms} can be explicitly obtained if we adopt conformal coordinates 
\be
{\Omega_{ab}\,\extd x^a\extd x^b}
=\Omega^2(|\zeta|) \extd\zeta \extd\bar{\zeta}.
\ee
The local conformal Killing vectors at the horizon then read 
\be
\eta^a\partial_{a}= \sum_k \eta_k (-\zeta^{k+1})\p_\zeta+  \sum_k \bar\eta_k (-\bar \zeta^{k+1})\p_{\bar\zeta}\, ,
\ee
with $\eta_k$ and $\bar\eta_k
$ constants, and the parameters of the spin-$s$ supertranslations can be expanded as
\be
\eta_{(s)}=\sum_{m,n} (\eta_{(s)})_{(m,n)} \zeta^m\bar\zeta^n.
\ee
The associated charges then expand accordingly 
\begin{align}
\sutr^{(s)}_{(m,n)} &= \int d\zeta d\bar\zeta\, \zeta^m\bar \zeta^n \sutr^{(s)}\\
 \suro_k=-\int d\zeta d\bar\zeta\, \zeta^{k+1 }\suro &, \quad
\bar \suro_k=-\int d\zeta d\bar\zeta\, \bar\zeta^{k+1 }\bar\suro\, , 
\end{align}
so that the algebra \eqref{eq:bms} in terms of these modes reads
\begin{subequations}
 \label{eq:4d-algebra}
\begin{align}
\{\suro_k,\, \sutr^{(s)}_{(m,n)}\} &= \left( \frac{s}{2}(k+1)-m\right)\, \sutr^{(s)}_{(k+m,n)} \\
\{\bar \suro_k,\, \sutr^{(s)}_{(m,n)}\} &= \left(\frac{s}{2}(k+1)-n\right)\, \sutr^{(s)}_{(m,k+n)} \\
\{\suro_n,\, \suro_m\} &= (n-m)\,\suro_{n+m}\, \\
\{\bar \suro_n,\, \bar  \suro_m\} &= (n-m)\,\bar 
\suro_{n+m}\, \\
\{\suro_n,\,\bar \suro_m\} &= 
\{\sutr^{(s)}_{(m,n)},\,\sutr^{(s)}_{(m',n')}\} = 0 \, . 
\end{align}
\end{subequations}

The subalgebra generated by $\suro_0,\suro_{\pm1}$ and $\bar \suro_0,\bar\suro_{\pm1}$ is the Lorentz algebra $sl(2)_L\,\oplus\,sl(2)_R\, \simeq so(3,1)$, which is the maximal finite subalgebra of the superrotation part. The ``spin-$s$ translations'' are spanned by the subset of $\mathcal{P}^{(s)}_{(m,n)}$, with $m,n=0,1,\dots,s$; and thus, the largest finite subalgebra of BMS$^{(s)}_4$ is given by the semidirect sum of Lorentz and spin-$s$ translations. This can be seen as follows. Let us determine the subset of supertranslations $\sutr_{(m,n)}^{(s)}$ that fall in the $(\frac{s}{2},\frac{s}{2})$ representation of $sl(2)_L\,\oplus\,sl(2)_R$. The left subalgebra is 
\begin{equation}
\big\{\suro_0,\suro_{\pm1}\big\}=\mp \suro_{\pm1} \qquad\qquad \big\{\suro_{+1},\suro_{-1}\big\}=2 \suro_0\,.
\end{equation}
The operators $\suro_{\pm 1}$ are the ladder operators (up to normalization) while $\suro_0$ measures the spin. The generator $\sutr_{(0,n)}^{(s)}$ corresponds to the highest weight state, and has spin $s/2$, since
\begin{equation}
\big\{\suro_{-1},\sutr_{(0,n)}^{(s)}\big\}=0\qquad\qquad \big\{\suro_{0},\sutr_{(0,n)}^{(s)}\big\}=\frac{s}{2}\,. 
\end{equation} 
The lowest weight one $\sutr_{(p,n)}^{(s)}$ is reached by acting with ${\cal J}_{+1}^p$ on the highest weight state, where $p>0$ solves $s-p=0$.  One notes that $\sutr^{(s)}_{(m,n)}$ with negative $m$ is never in a finite representation of the $sl(2)$ algebra, independently of the value of $s$.  The same considerations apply to the right sector. 

Thus, since the generators $\sutr_{(p,n)}^{(s)}$, with $p,n=0,\dots,s$, fall in the representation $(s/2,s/2)$ of the Lorentz algebra in four dimensions, which is of spin $s$, we {naturally} call them spin-$s$ translations; while spin-$s$ supertranslations correspond to the whole set of $\sutr_{(p,n)}^{(s)}$ with $p,n$ integers.  

Indeed, for $s=1$, the algebra \eqref{eq:4d-algebra} corresponds to the extension of the original BMS$_4$  \cite{Bondi:1962,Sachs:1962}, that includes superrotations \cite{Barnich:2011mi}. The maximal finite subalgebra is the Poincar\'e algebra, where the translations $\sutr_{(0,0)}^{(1)},\sutr_{(0,1)}^{(1)}, \sutr_{(1,0)}^{(1)}$ and $\sutr_{(1,1)}^{(1)}$ are expressed in the $(1/2,1/2)$ representation of the Lorentz algebra. 

In the case of $s=0$, the algebra \eqref{eq:4d-algebra} contains ```scalar supertranslations'' and it has been studied in \cite{Donnay:2015abr}. The maximal finite subalgebra is then given by $so(3,1)\oplus \mathbb{R}$.

For generic $s$, the algebra \eqref{eq:4d-algebra} corresponds to the algebraic deformation of BMS$_4$ given by $W(-s/2,-s/2;-s/2,-s/2)$, cf.~Eq.~(3.16) in \cite{Safari:2019zmc}.

Note that higher spin extensions of supertranslations (and superrotations) in $D=4$ were recently discussed in \cite{Campoleoni:2017mbt} in the context of the asymptotic behavior of massless higher spin fields on Minkowski space at null infinity, and applied to generalizations of Weinberg's soft theorems along the lines of Strominger \cite{Strominger:2013jfa}. In our context, it is amusing to show that higher spin symmetries arise near the horizon without the need of higher spin fields.

\subsection{Higher dimensions} The near horizon symmetry algebra BMS$^{(s)}_D$ in \eqref{eq:bms} has not been previously reported. 
As it was mentioned, if the spacelike section of the horizon metric is assumed to be conformal to the sphere $S^{D-2}$, the parameters $\eta^{a}$ must solve the conformal Killing equation 
\be
\nabla_a\eta_b+\nabla_b\eta_a=\frac{2}{D-2}\gamma_{ab}\nabla_c\eta^c
\ee
where $\gamma_{ab}$ is the metric on the round $S^{D-2}$ and $\nabla_a$ is the covariant derivative with respect to $\gamma_{ab}$. Requiring $\eta^{a}$ to be globally well-defined breaks the full set of diffeomorphisms on $S^{D-2}$, so that the associated generators $\mathcal{J}_a$ span the Euclidean conformal algebra in $D-2$ dimensions, $so(D-1,1)$, being equivalent to the $D$-dimensional Lorentz algebra. Therefore, in this case, the algebra reduces to the semi-direct sum of the Lorentz algebra and the spin-$s$ supertranslation part spanned by $\mathcal{P}^{(s)}$.

For $s=1$, the transformation laws in \eqref{eq:nh9}, with $r=1/(D-2)$, reduce to those of BMS$_D$ (see e.g. \cite{Barnich:2011ct}), so that we have obtained the first explicit realization of the BMS algebra in $D$ spacetime dimensions as near horizon symmetries. The BMS$_D$ algebra has been previously discussed  \cite{Tanabe:2011es,Kapec:2015vwa,Avery:2015gxa,Hollands:2016oma} at null infinity. In this case, the maximal finite subalgebra is the Poincar\'e algebra $iso(D-1,1)$, for which the translations clearly have spin 1.

An alternative way to see this comes from the scaling properties of the higher spin supertranslations with respect to the conformal sphere at the horizon. Note that according to Eq.\eqref{eq:nh4}, $\sutr$ is proportional to the $(D-2)$-dimensional volume element, and hence, it is a scalar density with scaling dimension $D-2$ under the $(D-2)$-dimensional Euclidean conformal algebra $so(D-1,1)$. Thus, the scaling dimension of $\sutr^{(s)}$ is given by $(r+1)(D-2)$, \emph{cf}.~Eq.~\eqref{eq:nh8}. Therefore, since the generators of the higher spin supertranslations in \eqref{eq:nh8} involve the integral over the $(D-2)$-dimensional spacelike section of the horizon, they possess scaling dimension $s$ provided that $(r+1)(D-2)-(D-2)=s$, i.e. for $r=s/(D-2)$, which  reduces to the  expected result for $s=1$.

\section{Near Horizon charges for Kerr--Taub--NUT }\label{Appen-2}

\newcommand{\kerr}{a}

The Kerr--Taub--NUT metric in Boyer--Lindquist coordinates \cite{Demianski:1966, Miller:1973, Griffiths:2009dfa}
\begin{align}
\extd s^2 = -\frac{\Delta}{\Sigma}
\big(\extd \hat{t}-(a \sin^2\theta -2n \cos\theta) \extd \phi \big)^2 +\frac{\Sigma}{\Delta}\extd r^2
+\Sigma \extd \theta^2 +\frac{\sin^2\theta}{\Sigma}\big(a \extd \hat{t}-(r^2+a^2+n^2)\extd \phi \big)^2 
\label{eq:kerrTaubNut1}
\end{align}
with
\eq{
\Delta := r^2-2Mr+\kerr^2-n^2, \qquad
\Sigma := r^2+(n+\kerr\cos\theta)^2
}{eq:kerr2}
has Killing horizons located at the zeros of $\Delta$, given by
\eq{
r_{\pm}=M\pm\sqrt{M^2-\kerr^2+n^2}\,.
}{eq:kerr3}
The mass $M$, angular momentum $J$, and the rotation parameter $\kerr$ can be expressed in terms of outer and inner horizon radii $r_\pm$ and the NUT charge $n$,
\begin{align}
M &=\frac{r_++r_-}{2}\,,\quad J=Ma\,,\quad 
\kerr^2 = n^2+r_+r_- \,,
\label{eq:kerr4c}
\end{align}
and the surface gravity is given by
\begin{equation}
    \hat{\kappa}=\frac{r_{+}-r_{-}}{2\left(2n^{2}+r_{+}\left(r_{+}+r_{-}\right)\right)}.
\end{equation}
The following coordinate change 
\begin{align}\label{eq:kerr8}
r & =r_+ + \frac{r_+-r_-}{4\Sigma_+}\,\rho^2\,, \qquad 
\phi  =\varphi+\frac{2\hat{\kappa} a}{r_{+}-r_{-}}\hat{t}
\end{align}
with \begin{equation}
\Sigma_+ :=r_+^2+(n+a\cos\theta)^2
\end{equation}
shifts the Kerr black hole horizon at $r=r_+$ to $\rho=0$, and brings the metric into a form that fits our near horizon expansion in \eqref{eq:nh1}.

In order to fulfill the Heisenberg-like boundary conditions it is necessary to rescale the time coordinate as 
\begin{equation}
    \hat{t} = \frac{\kappa_{\text{\tiny{H}}}}{\hat{\kappa}}t,
\end{equation}
where $\kappa_{\text{\tiny{H}}}$ stands for an arbitrary constant without variation ($\delta \kappa_{\text{\tiny{H}}}=0$), so that the near horizon expansion reads
\begin{subequations}
\label{eq:kerreqs}
\begin{align}
g_{tt} & = -\kappa_{\text{\tiny{H}}}^2 \rho^2 +{\cal O}\left(\rho^{4}\right) \\
g_{t\rho}& =g_{t\theta}=0 \\
g_{t\varphi}  &= \frac{\kappa_{\text{\tiny{H}}}}{2\Sigma_{+}^{2}}\left[2ar_{+}\sin^{2}\theta\left(2n^{2}+r_{+}^{2}+r_{-}r_{+}\right)
+\left(r_{+}-r_{-}\right)\left(a\sin^{2}\theta-2n\cos\theta\right)\Sigma_{+}\right]\rho^{2}+O\left(\rho^{4}\right)
\\
g_{\rho\rho} & =1+{\cal O}\left(\rho^{2}\right)\\
g_{\rho\theta} & =\frac{a (n+a\cos\theta)\sin\theta}{\Sigma_+}\,\rho+{\cal O}\left(\rho^{3}\right)\\
g_{\rho\varphi} & =g_{\theta\varphi}=0\\
g_{\theta\theta} & =\Sigma_++O\left(\rho^{2}\right)\\
g_{\varphi\varphi} & =\frac{\left(2n^2+r_+(r_++r_-) \right)^{2}\sin^{2}\theta }{\Sigma_+} + {\cal O}\left(\rho^{2}\right)\,.
\end{align}
\end{subequations}

In particular, the horizon metric $\Omega_{ab}$ is given by
\begin{equation}
\Omega_{ab}\,\extd x^a\extd x^b = \Sigma_+\,\extd\theta^2 
+\frac{\left(2n^2+r_+(r_++r_-) \right)^{2}\sin^{2}\theta }{\Sigma_+}\,\extd\varphi^2\,.
\label{eq:kerr6}
\end{equation}
{The above horizon metric  is topologically a 2-sphere, albeit not a round one and in general the South and North poles $\theta=0,\pi$ points have conical singularities. To verify the latter one may study the geometry near the poles. Near the poles, $\theta=0+{\cal R}$ or $\theta=\pi-{\cal R}$, \eqref{eq:kerr6} to leading order in ${\cal R}$ takes the form
\begin{align}\label{horizon-near-poles}
\!\!\!\!\!\Omega_{ab}\,\extd x^a\extd x^b \sim \Sigma^\sigma_+\left[\,\extd {\cal R}^2
+\left(1-\sigma\frac{2na}{\Sigma^\sigma_+} \right)^{2}{\cal R}^2\extd\varphi^2\right]
\end{align}
where $\Sigma^\sigma_+=r_+^2+(n +\sigma a)^2$ and $\sigma=\pm 1$ (denoting the North and  South poles). As we see, due to the existence of the NUT charge $n$, there is 
\begin{equation}\label{deficit-angle}
\Delta_\sigma\equiv 2\pi\sigma \frac{2na}{\Sigma^\sigma_+}
\end{equation}
deficit/excess angle. To examine the topology, we compute the Euler character. Due to existence of conical deficits, 
\begin{equation}\label{Euler-deficit}
\chi=\frac{1}{4\pi}\int \extd\theta\extd\varphi \sqrt{\Omega}\,R_{\textrm{\tiny H}} +\frac{1}{2\pi}(\Delta_++\Delta_-)
\end{equation}
where $R_{\textrm{\tiny H}}$ is the Ricci scalar of the horizon,
\begin{align}
R_{\textrm{\tiny H}}&= \frac{1}
{\Sigma_+^3}\big[-2 r_+^4+r_+^3 r_-+3 r_+^2 r_-^2+  {-3 r_+^2 n^2+15 r_+ r_- n^2+}{10 n^4+15 a^3 n \cos \theta+} \nonumber \\
&  +3 a^2 \left(r_+ (r_++r_-)+2 n^2\right) \cos(2 \theta )+a^3 n \cos(3 \theta )\big].
\label{kerr11}
\end{align}
Then,
\begin{equation}
\begin{split}
\frac{1}{4\pi}\int R_{\textrm{\tiny H}}&=\frac{2(2n^2+r_+(r_++r_-))^2}{r_+^2(4n^2+(r_++r_-)^2)}\cr
\frac{1}{2\pi}\left(\Delta_++\Delta_-\right)&=\frac{2na}{r_+^2+(n+a)^2}-\frac{2na}{r_+^2+(n-a)^2}
\end{split}
\end{equation}
With the above, and using \eqref{eq:kerr4c} one can readily show that $\chi=2$. 
}


From the near horizon expansion \eqref{eq:kerreqs} one can read the generators $\sutr$ and $\suro_a^\tH$ defined in \eqref{eq:nh4} and \eqref{eq:angelinajolie},
\begin{align}
    \sutr &= \frac{2 n^2+r_+(r_++r_-)}{8\pi G}\,\sin\theta \label{eq:kerr7}\\
    \suro_a^\tH &=\frac{\delta_{a}^{\varphi}}{2\Sigma_{+}^{2}}\left[\left(r_{+}-r_{-}\right)\left(2n\cos\theta-a\sin^{2}\theta\right)\Sigma_{+}
 -2ar_{+}\left(2n^{2}+r_{+}\left(r_{+}+r_{-}\right)\right)\sin^{2}\theta\right]
    \,. 
\end{align}
Setting the NUT charge to zero, $n=0$, yields the results quoted in the main text \eqref{eq:kerr}. 
Note that for the Kerr--Taub--NUT black hole, the ``field strength'' in \eqref{eq:nh11b} is given by
\begin{align}\label{eq:kerr9}
 {F}_{\varphi\theta} &= -\partial_\theta\suro_\varphi^\tH  =\frac{\left(r_+ (r_++r_-)+2 n^2\right) \sin\theta }
{4 \Sigma_+^3} 
\big[4 r_+^3 n + 18 r_+^2 r_- n - 6 r_+ r_-^2 n + 26 r_+ n^3 - 10 r_- n^3\nonumber\\
&\!\!\!\!\!\! +3 a \left(r_+ \left(4 r_+^2 + 3 r_+ r_- - r_-^2\right)+(11 r_+ - 5 r_-) n^2\right) \cos\theta 
+6 a^2 (r_+ - r_-) n \cos(2 \theta) - a^3 (r_++r_-) \cos(3 \theta )\big].
\end{align}

{Noting $F=\extd\suro^\tH$ (as a two-form on the horizon) we have
\begin{align}\label{eq:kerr10}    
\int_{H}\!{ F} = \int_{H_{\cal R}} F +( {\cal F}_+-{\cal F}_-)
\end{align}
where  $H_{\cal R}$ denotes a part of the horizon geometry in which the near pole region \eqref{horizon-near-poles} has been excised in the ${\cal R}\to 0$ limit and ${\cal F}_\sigma$ appears due to the deficit/excess angles in poles:
\eq{
{\cal F}_\sigma=\oint_{{\cal R}} \suro^\tH\big|_{\sigma}=  {2}{n\hat\kappa\sigma}(\Delta_\sigma),}{eq:DeltaJ}
where the integral is over the ``complement part''  of the near pole geometry \eqref{horizon-near-poles} in which $(1-\frac{\Delta_\sigma}{2\pi})\varphi$ ranges from $0$ to $\Delta_\sigma$, 
with $\Delta_\sigma$  given in \eqref{deficit-angle}.
A direct computation using \eqref{eq:kerr9} yields
$$
\int_{H_{\cal R}} F=n\hat\kappa \int\extd^2x\sqrt{\Omega}\,R_{\textrm{\tiny H}},
$$
and therefore, recalling \eqref{Euler-deficit}, we have
\eq{
\frac{1}{4\pi\hat\kappa}\int_{H}F =n \chi. 
}{eq:kerr11} 
The above is compatible 
with  quantization of the NUT charge.
}

%

\end{document}